\documentclass[prl,twocolumn]{revtex4}
\usepackage{graphicx}
\usepackage{amsmath,amssymb}

\begin{document}
\title{Orbital-free extension to Kohn-Sham density functional theory equation of state calculations: application to silicon dioxide}
\author{Travis Sjostrom and Scott Crockett}
\affiliation{Theoretical Division, Los Alamos National Laboratory,
Los Alamos, New Mexico 87545}
\date{July 16, 2015}
\begin{abstract}
The liquid regime equation of state of silicon dioxide SiO$_2$ is calculated via quantum molecular dynamics in the density range 5 to 15 g/cc and with temperatures from 0.5 to 100 eV, including the $\alpha$-quartz and stishovite phase Hugoniot curves. 
Below 8 eV calculations are based on Kohn-Sham density functional theory (DFT), above 8 eV a new orbital-free DFT formulation, presented here, based on matching Kohn-Sham DFT calculations is employed.
Recent experimental shock data is found to be in very good agreement with the current results. Finally both experimental and simulation data are used in constructing a new liquid regime equation of state table for SiO$_2$.
\end{abstract}
\maketitle
%
Access to the accurate equation of state (EOS) of materials over wide ranges in density and temperature, and in particular for matter in extreme conditions of significantly elevated temperature and density with respect to ambient conditions, is important in various arenas. 
Some systems of interest include dense astrophysical plasmas as exist in the interiors of giant planets, as well as warm dense matter, which is increasingly studied in high energy density laboratory experiments, and also in development of inertial confinement fusion \cite{Drake}.
A particular case, presented here is that of quartz (SiO$_2$), a ubiquitous mineral in Earth's composition, it is also constituent in exoplanet modeling at higher temperatures and pressures than at inner Earth conditions \cite{Millotetal}.
This material is also important as a window material in high compression shock experiments, typically in the quartz liquid regime above 100 GPa and 5000 K \cite{KD09,KD13}. 
Hence, an accurate EOS for quartz is critical for determining properties of other materials through shock experiments. 

In general these extreme conditions represent a significant and current challenge of high energy density physics. Experimental results remain sparse, and from a theoretical standpoint the ions exhibit moderate to strong coupling while the electrons require quantum treatment \cite{IPAM}. 
This then necessitates numeric simulations for which quantum molecular dynamics (QMD) based on Kohn-Sham density functional theory (DFT) has emerged as the state of the art. Kohn-Sham DFT, however, becomes computationally prohibitive with increasing temperature as the number of required orbitals increases with temperature and in general the method scales as the cube of the number of orbitals. Though material and density dependent, Kohn-Sham calculations are, thus, generally viable below 10 eV or so. In this work we develop an orbital-free DFT formulation to combine with Kohn-Sham results and extend QMD simulations to very high temperatures.

In QMD, the ions are treated classically and moved according to Newton's equations, where the force on each ion is found from the Coulomb repulsion between all ions, and from the neutralizing electron charge density. The electron density, $n$, is found at each ionic configuration by DFT. This is done by minimizing the free energy, which is given by the density functional \cite{ParrYang}
\begin{equation}
  F[n]=F_s[n] + F_H[n] + F_{xc}[n] + F_{ei}[n] 
  \label{eq:fe1}
\end{equation}
where $F_s$ is the non-interacting free energy comprised of both kinetic and entropic parts, $F_H$ is the Hartree energy or direct Coulomb interaction between the electrons, $F_{\rm ei}$ is the electron-ion Coulomb interaction, and $F_{xc}$ is defined as the remainder of the total free energy, which includes the quantum mechanical exchange and correlation as well as the excess kinetic and entropic terms. Of the contributions neither $F_s$ nor $F_{xc}$ have exact formulations in terms of the density alone. Given the same orbital-free $F_{xc}$ approximation, the only difference in approach of orbital-free DFT from Kohn-Sham DFT is that the non-interacting free energy, $F_s$, is found from an approximate density functional instead of being exactly obtained through the calculation of single particle orbitals. 

In recent years the orbital-free approach at finite temperature has gained attention, with most results being for hot dense systems where the Thomas-Fermi approximation is employed for $F_s$ \cite{Lambert06}. Various works have offered density gradient corrections to Thomas-Fermi that improves results moderately \cite{Perrot79,Daneletal,ftgga,vt84f}. None of these functionals, though,  have reached the accuracy of Kohn-Sham across temperature regimes. A recent nonlocal functional has been shown to be highly accurate across temperature and density regimes \cite{SjostromDaligault}, but is requisite on Kohn-Sham derived pseudopotentials, which may not be as transferable. Subsequently an accurate and general orbital-free funcitonal remains elusive.

 Previous works \cite{sheppardetal,wangetal} have attempted to connect high temperature Thomas-Fermi calculations with low temperature Kohn-Sham results, for low atomic number systems. In this work we develop and implement a method to obtain a simple and accurate $F_s$ beyond Thomas-Fermi, applicable to a wide range of materials and densities, for warm to hot systems, which smoothly extends Kohn-Sham results beyond the lower temperature region that is currently computationally accessible.
Then in an application to SiO$_2$ a new liquid phase EOS is developed utilizing the current QMD calculations and compared with experimental results. 

We first introduce an orbital-free $F_s$ of the following form
\begin{align}
  F_s[n] = F_{TF}[n] + \lambda F_{vW}[n]
  \label{eq:fs}
\end{align}
with
\begin{align}
  F_{TF}[n] &= \int F^{HEG}(n(\mathbf{r}),T) / V d\mathbf{r}\;, \\
  F_{vW}[n] &= \int \frac{|\nabla n(\mathbf{r})|^2}{8 n(\mathbf{r})} d\mathbf{r}\;.
\end{align}
Here $F_{TF}$ is the Thomas-Fermi approximation which takes for the noninteracting free energy that of the homogeneous electron gas, $F^{HEG}$, with density equal to that of the local density and at the system temperature $T$; $F_{vW}$ is the von-Weizsacker gradient correction which has no explicit temperature dependence. The von Weizsacker coefficient $\lambda$, is material and density dependent, and is to be determined through matching conditions between orbital-free and Kohn-Sham calculations at a specific density. In a more general way one could write $\lambda(n_0)$, with $n_0$ being the average electron density, here though, $\lambda$ is treated as a constant and in that way Eq. (\ref{eq:fs}) represents a best approximation at a given density to the more general $F_S$, which is a single functional over varying densities.

\begin{figure}
  \centering
  \includegraphics[width=3.0in]{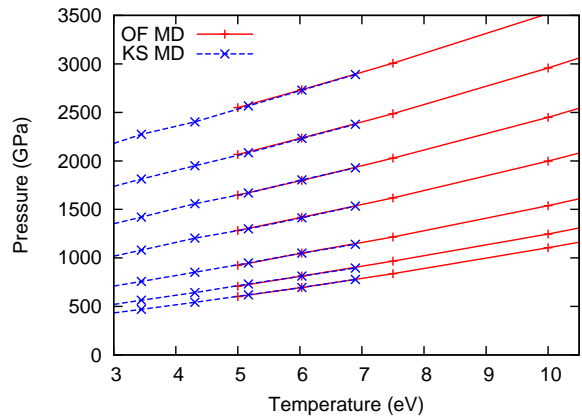}
  \caption{Pressure results of Kohn-Sham (KS) MD and orbital-free (OF) MD near the match temperature of 6 eV, for isochores of 11, 10, 9, 8, 6.9, 6.43, and 5.57 g/cc, $\lambda$ ranges from 0.18-0.22.}
  \label{fig:match}
\end{figure}

The procedure to determine $\lambda$ is straight forward. First we evaluate the pressure along an isochore using Kohn-Sham MD, until the temperature is high enough that the calculations becomes intractable (which of course depends on computational resources) but should be at least 5 to 10 eV. Below which the orbital-free approach will become inaccurate due to issues such as molecular bonding. Then at a given match temperature a few orbital-free calculations are performed with initial guesses for $\lambda$ and the $\lambda$ is determined which reproduces the Kohn-Sham data in pressure, $\lambda$ is then fixed for that density. Figure \ref{fig:match} shows results of the matching method over a range of densities, a strong argument is made for the approach in that even though the match for pressure is made at the single point $T=6$ eV, the change with temperature, or slope in Fig. \ref{fig:match}, is in near exact agreement between the two DFT methods. Also at very high temperature $\sim$100 eV, our results come into agreement with the Thomas-Fermi based MD, which is correct in the high-$T$ limit.

The details of the calculations are as follows. The Kohn-Sham SiO$_2$ calculations were performed in the liquid regime above 5.5 g/cc and 0.5 eV using the Quantum-\textsc{Espresso} program \cite{qespresso}. We included 72 total atoms in the calculations and all were performed with the $\Gamma$-point only. Some calculations were performed with a 2x2x2 $k$-point grid, without noticeable differences to the Gamma-point calculations. PAW pseudopotentials \cite{Blochl} were employed, along with a temperature dependent implementation of the AM05 exchange-correlation functional \cite{am05,ftggaxc}, and a plane wave energy cutoff of 50 Ry was used. The molecular dynamics were performed at constant temperature using the Anderson thermostat. In order to calculate the Hugoniot, described below, the initial crystalline states of $\alpha$-quartz and stishovite were calculated at the experimentally prescribed densities of 2.65 and 4.29 g/cc at room temperature, using the primitive cell of each phase with fully converged $k$-point grids.

In the orbital-free case, the same number of atoms were included as for the Kohn-Sham case, and the electron density was optimized on a regular 64$^3$ numeric grid. Local pseudopotentials were generated for each density and temperature, according to the prescription given in Ref. \onlinecite{Lambert06}, with a cutoff radius of 0.6 time the Wigner-Seitz radius. Since the orbital-free calculations were performed only above 5 eV, only the temperature dependent local density approximation (LDA) exchange-correlation \cite{ftldaxc} was used for simplicity. The orbital-free molecular dynamics were completed in the isokinetic ensemble \cite{Bussietal}.

The key application of the above QMD calculations is in construction and validation of a far reaching EOS valid at arbitrary densities and temperatures. While experimental data serves as the traditional constraining input for EOS construction, it is particularly lacking in the warm dense matter regime. Further this regime falls in the region of interpolation between the low and high temperature models upon which the EOS is built, precisely where constraining input is most needed. The current QMD calculations then provide that input.

The overall EOS model itself is a summation of three independent terms for the total Helmholtz free energy, given in terms of the material density $\rho$ and the temperature $T$,
\begin{align}
  F_{tot} = F_0(\rho) + F_{\mathrm{i}}(\rho,T) + F_{\mathrm{e}}(\rho,T)\;.
\end{align}
The contributions here are the cold curve, $F_0$, the ion thermal contribution $F_{\mathrm{i}}$, and the electron thermal contribution $F_{\mathrm{e}}$.  First of all the cold curve, representing zero temperature electrons and ions, is constructed from a modified Lennard-Jones model in the expanded region and Thomas-Fermi-Dirac theory at very high pressures, with a Birch-Murnaghan model in the interim.  Next the ion thermal portion, due to the thermal motion of the ions, is built from a Debye model for temperatures less than the melt temperature, and interpolating to an ideal gas at high temperatures, leaving a poorly described liquid-like warm dense matter state. Finally the electron thermal portion, which comprises the thermal excitation of electrons above the ground state, is evaluated via the Thomas-Fermi average atom model.

In order to effect an accurate EOS, the various parameters of the constituent models need to be constrained by input data from experiment or further theoretical calculations. 

In the case of SiO$_2$ there is shock data providing the pressure-density curve of the Hugoniot for $\alpha$-quartz \cite{KD09} and more recently for stishovite \cite{Millotetal}, as well as some recent shock release data for $\alpha$-quartz \cite{KD13} which provides some off-Hugoniot data. This data alone is inadequate for a full ranging EOS, but when combined with wide ranging QMD data a more complete view is established, and an accurate EOS may be determined.


The Hugoniot itself is calculated via the Rankine-Hugoniot jump conditions \cite{DuvallGraham}, which relates the equilibrium pre-shock and post-shock states by consideration of conservation laws,
\begin{align}
  E-E_0=(P+P_0)(V_0-V)/2\;.
\label{eq:hug}
\end{align}
Here $E$, $P$ and $V$ are the internal energy, pressure and volume respectively, and the 0 subscript denotes the initial state. For $\alpha$-quartz and stishovite the initial conditions are densities of 2.65 g/cc and 4.29 g/cc respectively, and at ambient pressure and temperature. The initial internal energy is calculated via Kohn-Sham DFT and then using Kohn-Sham MD, an isochore or isotherm is followed until conditions are such that  Eq. (\ref{eq:hug}) is satisfied. However when extending to orbital-free MD there is a shift in the internal energy due to change in the pseudopotential and the functional. As part of the matching scheme this shift is found at the match point temperature and density, so that the initial state energy calculated by Kohn-Sham DFT can be used. Additionally since there is a different pseudopotential used at each temperature and density with the orbital-free DFT, a second shift in energy must be accounted for, which is found by performing two orbital-free calculations at the same temperature and density but with different pseudopotentials.
\begin{figure}
  \centering
  \includegraphics[width=3.0in]{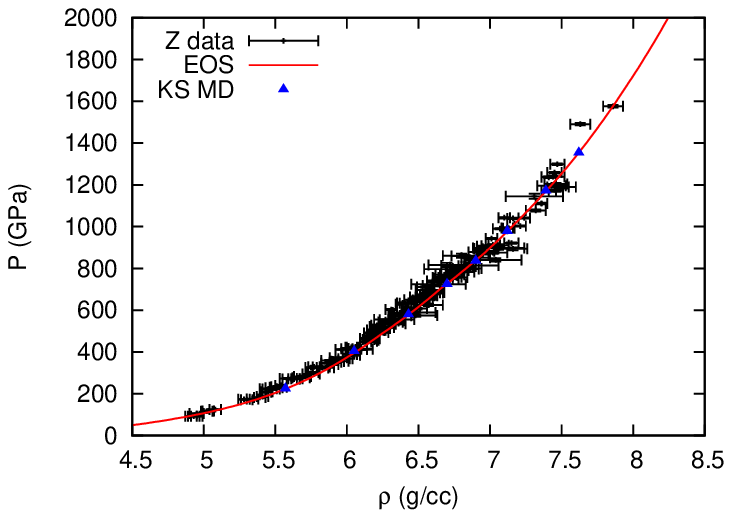}
  \includegraphics[width=3.0in]{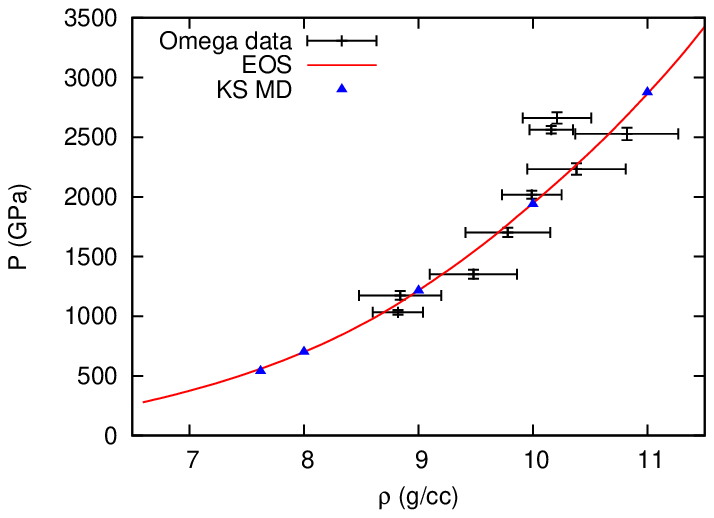}
  \includegraphics[width=3.0in]{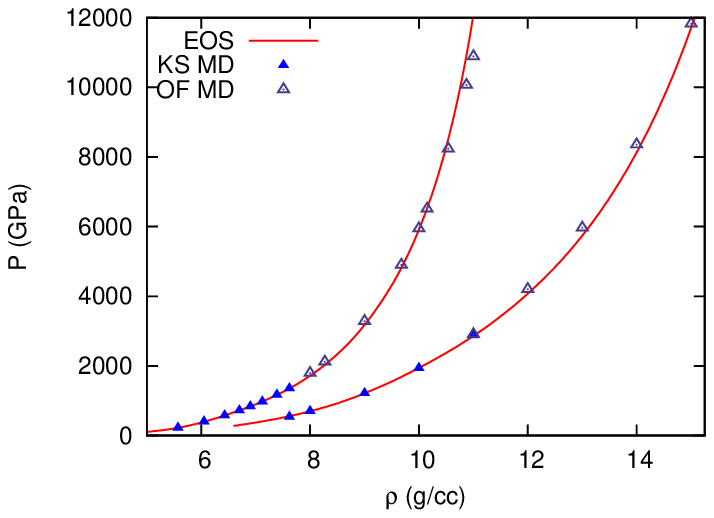}
  \caption{Comparison of experimental results, Refs. \onlinecite{KD09,KD13,Millotetal}, with current QMD and EOS shock Hugoniot results. Upper panel shows $\alpha$-quartz, middle panel shows stishovite, and the lower panel shows $\alpha$-quartz on the left curve and stishovite to the right.}
  \label{fig:hug}
\end{figure}

The resulting Hugoniots for $\alpha$-quartz and stishovite are shown in Fig. \ref{fig:hug}. The very good agreement between experiment and QMD, provides validity to the QMD results for extension to regions where there is no experimental data, such as the high pressure Hugoniot extensions shown in the lower panel of Fig. \ref{fig:hug}. A benefit of the QMD calculation is that the temperature is also calculated along the Hugoniot. This is not always available from the shock data, as it is not here with the $\alpha$-quartz data. Along the $\alpha$-quartz Hugoniot we find the temperature increases from 0.86 eV at 5.57 g/cc to 6.73 eV at 7.62 g/cc to 23.34 eV at 10.00 g/cc.

\begin{figure}
  \centering
  \includegraphics[width=3.0in]{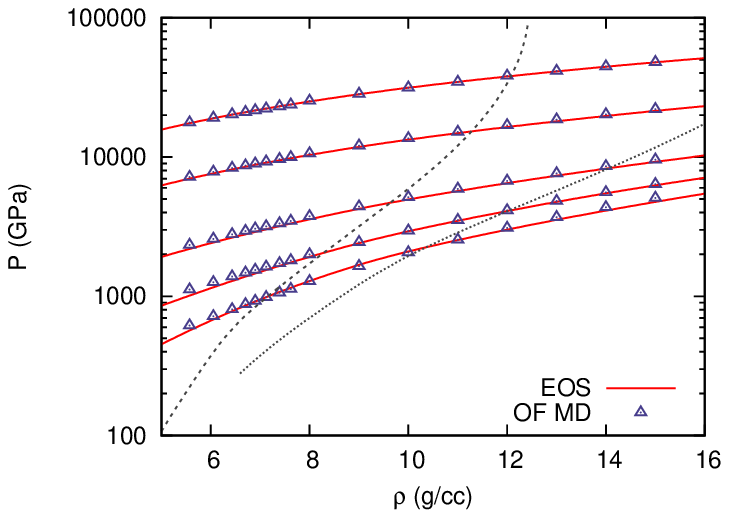}
  \includegraphics[width=3.0in]{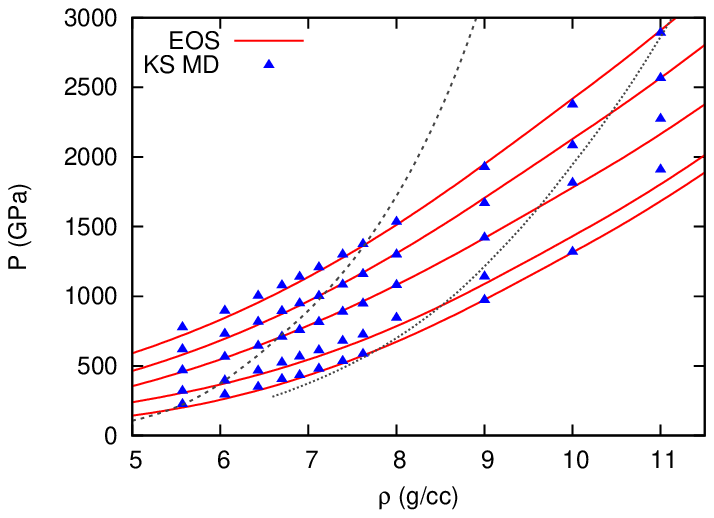}
  \caption{Comparison of pressures between the QMD results and the current EOS for liquid SiO$_2$ isotherms. From top to bottom temperatures are 100, 50, 20, 10 ,5 eV in the upper panel and 6.89, 5.17, 3.45, 1.72, 0.86 eV in the lower panel. The $\alpha$-quartz (dashed) and stishovite (dotted) Hugoniots calculated from the EOS are shown for reference.}
  \label{fig:isotherm}
\end{figure}

Away from the Hugoniot  the broad range of QMD results can be compared directly with the EOS. A subset of QMD calculations is plotted in Fig. \ref{fig:isotherm} showing the range of temperatures for 0.86-100 eV and densities from 5-15 g/cc. The higher temperature orbital-free results were instrumental in constraining the ion thermal portion of the EOS through the higher temperature liquid phase approaching the Thomas-Fermi-Dirac limit. 

\begin{figure}
  \centering
  \includegraphics[width=3.0in]{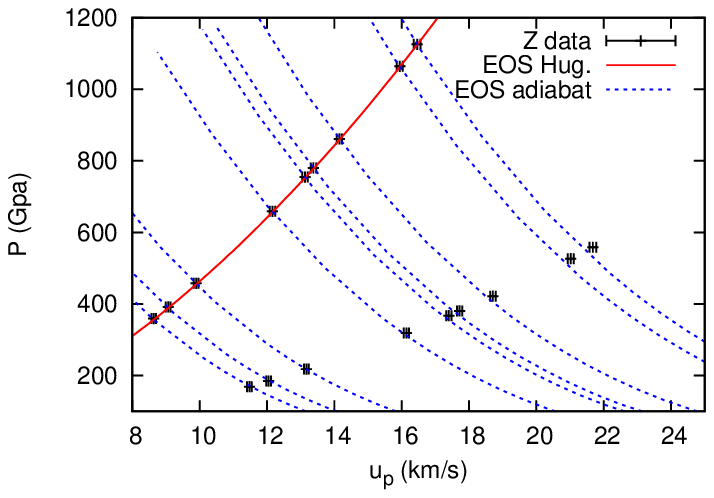}
  \includegraphics[width=3.0in]{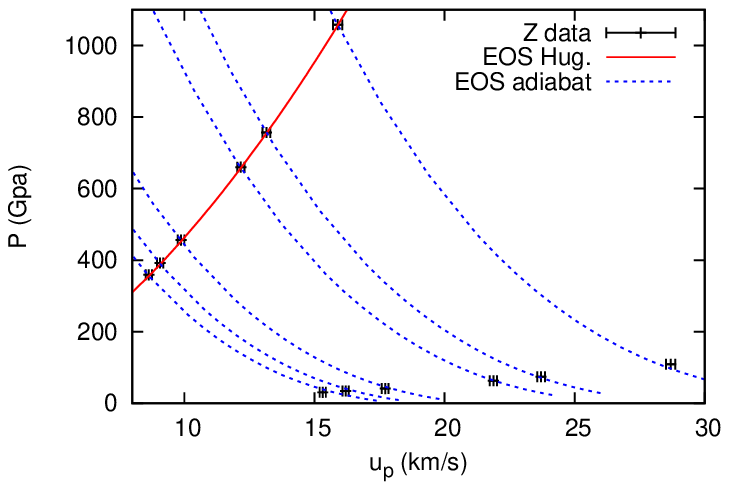}
  \includegraphics[width=3.0in]{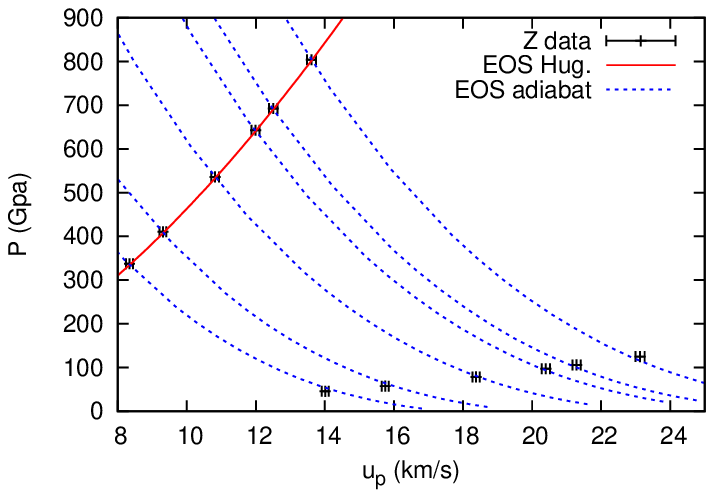}
  \caption{Good agreement is seen between the experimental Z-machine shock release data, Ref. \onlinecite{KD13}, and the EOS calculated adiabats.}
  \label{fig:release}
\end{figure}

Finally we show the agreement of the resulting EOS with the shock release data in Fig. \ref{fig:release}. Here the adiabats are calculated within the EOS which pass through the experimental release points along the Hugoniot. Good agreement is shown with experiment for the release in each of the three materials, which also with the isotherm data of Fig. \ref{fig:isotherm} demonstrate high accuracy for the EOS away from the Hugoniot.


In summary we have developed a robust and prescriptive method for accurately extending Kohn-Sham DFT based molecular dynamics simulations by orbital-free DFT simulations, for equation of state calculations including calculation of the Hugoniot to very high temperatures. This is important as the Kohn-Sham MD calculations have been shown to accurately characterize both solid and liquid systems and have become a gold standard. Yet due to temperature scaling issues, going beyond 10 eV or so is formidable and despite advances in computing resources seems poised to remain so. The current orbital-free extension alleviates this bottle neck, allowing for accurate results in conjunction with the Kohn-Sham method, from zero temperature through the high temperature Thomas-Fermi limit.

As a relevant application we constructed a wide ranging EOS for SiO$_2$. The EOS constrained by the QMD calculations shows very good agreement with the recent shock experiment data. An immediate and critical application of the new EOS lies in design and analysis of high-pressure shock experiments of different materials where $\alpha$-quartz has been used extensively as a window material, and hence determination of these materials relies on a highly accurate SiO$_2$ EOS.

We would like to thank Eric Chisolm and Carl Greeff for helpful discussions during this research.
This work was carried out under the auspices of the National Nuclear Security Administration of the U.S. Department of Energy (DOE) at Los Alamos National Laboratory under Contract No. DE-AC52-06NA25396.

\end{document}